\documentclass[iop]{emulateapj}
\pdfoutput=1 
\usepackage{amsmath,amstext}
\usepackage[T1]{fontenc}
\usepackage{apjfonts}
\usepackage{hyperref}
\usepackage[figure,figure*]{hypcap}

\usepackage{color}

\newcommand{\maya}[1]{\textcolor{black}{#1}}

\begin{document}

\title{Does the Black Hole Merger Rate Evolve with Redshift?}
\author{Maya Fishbach}
\affiliation{Department of Astronomy and Astrophysics, University of Chicago, Chicago, IL 60637, USA}
\author{Daniel E. Holz}
\affiliation{Enrico Fermi Institute, Department of Physics, Department of Astronomy and Astrophysics,\\and Kavli Institute for Cosmological Physics, University of Chicago, Chicago, IL 60637, USA}
\affiliation{Kavli Institute for Particle Astrophysics \& Cosmology and Physics Department, Stanford University,\\ Stanford, CA 94305}
\author{Will M. Farr}
\affiliation{School of Physics and Astronomy and Birmingham Institute of Gravitational Wave Astronomy\\University of Birmingham,
Birmingham, B15 2TT, England,}
\affiliation{Center for Computational Astrophysics, Flatiron Institute\\  162 Fifth Avenue, New York NY 10010,
United States,}
\affiliation{Department of Physics and Astronomy, Stony Brook University,
Stony Brook NY 11794-3800, United States}

\begin{abstract}
We explore the ability of gravitational-wave detectors to extract the redshift distribution of binary black hole (BBH) mergers. The evolution of the merger rate across redshifts $0 < z \lesssim 1$ is directly tied to the formation and evolutionary processes, providing insight regarding the progenitor formation rate together with the distribution of time delays between formation and merger. Because the limiting distance to which BBHs are detected depends on the masses of the binary, the redshift distribution of detected binaries depends on their underlying mass distribution. We therefore consider the mass and redshift distributions simultaneously, and fit the merger rate density, ${dN}/{dm_1\,dm_2\,dz}$. Our constraints on the mass distribution agree with previously published results, including evidence for an upper mass cutoff at $\sim 40 \ M_\odot$. Additionally, we show that the current set of six BBH detections are consistent with a merger rate density that is uniform in comoving volume. Although our constraints on the redshift distribution are not yet tight enough to distinguish between BBH formation channels, we show that it will be possible to distinguish between different astrophysically motivated models of the merger rate evolution with $\sim 100$--$300$ LIGO-Virgo detections (to be expected within 2--5 years). Specifically, we will be able to infer whether the formation rate peaks at higher or lower redshifts than the star formation rate, or the typical time delay between formation and merger. Meanwhile, with $\sim 100$ detections, the inferred redshift distribution will place constraints on more exotic scenarios such as modified gravity.
\end{abstract}

\section{Introduction}
The redshift dependence of the binary black hole (BBH) merger rate carries information about the processes by which BBHs evolve and merge, including the environments in which they form, the star formation rate (SFR), and the time-delay distribution. By measuring the luminosity distance\footnote{Throughout we fix the cosmological parameters to their Planck 2015 values~\citep{PlanckCosmology} to convert between the GW-measured luminosity distance and the cosmological redshift of the source. Changes to these parameters within the current range of uncertainties will not have any qualitative impact on our conclusions.} to detected sources, the current generation of ground-based gravitational-wave (GW) detectors will be able to measure the redshift distribution of BBH mergers up to redshifts $z \sim 1$ \citep{2016ApJ...818L..22A,Abbott:OS}. The inferred redshift distribution will provide important clues regarding the BBH formation channel. For example, in the classical isolated binary evolution channel, the redshift evolution follows the SFR convolved with a distribution of time delays between formation and merger \citep{2015ApJ...806..263D,2016ApJ...818L..22A,2016Natur.534..512B}. Meanwhile, in the dynamical formation channel, the evolution of BBH mergers is tied to the evolution of globular clusters \citep{2041-8205-836-2-L26,Rodriguez:2018}. If BBHs are primordial, they are expected to largely follow the dark matter distribution \citep{2016PhRvL.117t1102M,PhysRevLett.119.221104}. Furthermore, several exotic scenarios, such as gravitational leakage~\citep{DGP,2007ApJ...668L.143D,2018arXiv180108160P} or a significant population of strongly lensed BBH systems~\citep{2018arXiv180205273B,2018arXiv180307851S}, would leave an imprint on the inferred redshift distribution.

In this work we consider the BBH merger rate density as a function of the component masses, $m_1$ and $m_2$, and redshift, $z$. The mass and redshift distributions must be fit simultaneously, because the detection efficiency of GW detectors depends on the component masses as well as the distance to the source, as discussed in \S\ref{sec:zdet}. We parametrize the mass distribution as a power law with a variable upper mass cutoff, as in ~\cite{2017ApJ...851L..25F}.
For the redshift distribution, we consider two parametrizations. The first parametrization is motivated by the low-redshift SFR, and assumes that the BBH merger rate follows the comoving volume to \maya{zeroth order in redshift}. This model can fit astrophysically motivated redshift distributions, including metallicity-weighted SFRs convolved with various time-delay distributions. \maya{The second model allows for much more extreme deviations from a uniform in comoving volume merger rate, even at low redshifts,} making it less suitable for distinguishing between different formation channels but more sensitive to the exotic scenarios discussed above, including modified gravity. These two models are described in \S\ref{sec:model}.

In \S\ref{sec:detections} we fit a joint mass-redshift distribution to the first six BBH detections announced by the LIGO-Virgo Collaboration (LVC): GW150914, LVT151012, GW151226, GW170104, GW170814, and GW170608 \citep{2016PhRvL.116f1102A,2016PhRvD..93l2003A,PhysRevLett.116.241103,2017PhRvL.118v1101A,PhysRevLett.119.141101,2017ApJ...851L..35A}. For simplicity we treat LVT151012, which has an 87\% probability of having astrophysical origin, as a full detection \citep{2016PhRvX...6d1015A}.
\maya{Although it has recently been suggested otherwise \citep{2018arXiv180205273B},} we find that this set of detections is entirely consistent with a redshift distribution that is uniform in comoving volume. We show how future detections will improve the measurement of the BBH merger rate as a function of component masses and redshift in \S\ref{sec:future}. We predict that with a few hundred BBH detections by LIGO-Virgo operating at design sensitivity, the measurement of the BBH redshift distribution will be precise enough to distinguish between different formation channels.

\section{Detected redshift distribution}
\label{sec:zdet}
\begin{figure}
\label{fig:cumpzdet}
\includegraphics[width=0.5\textwidth]{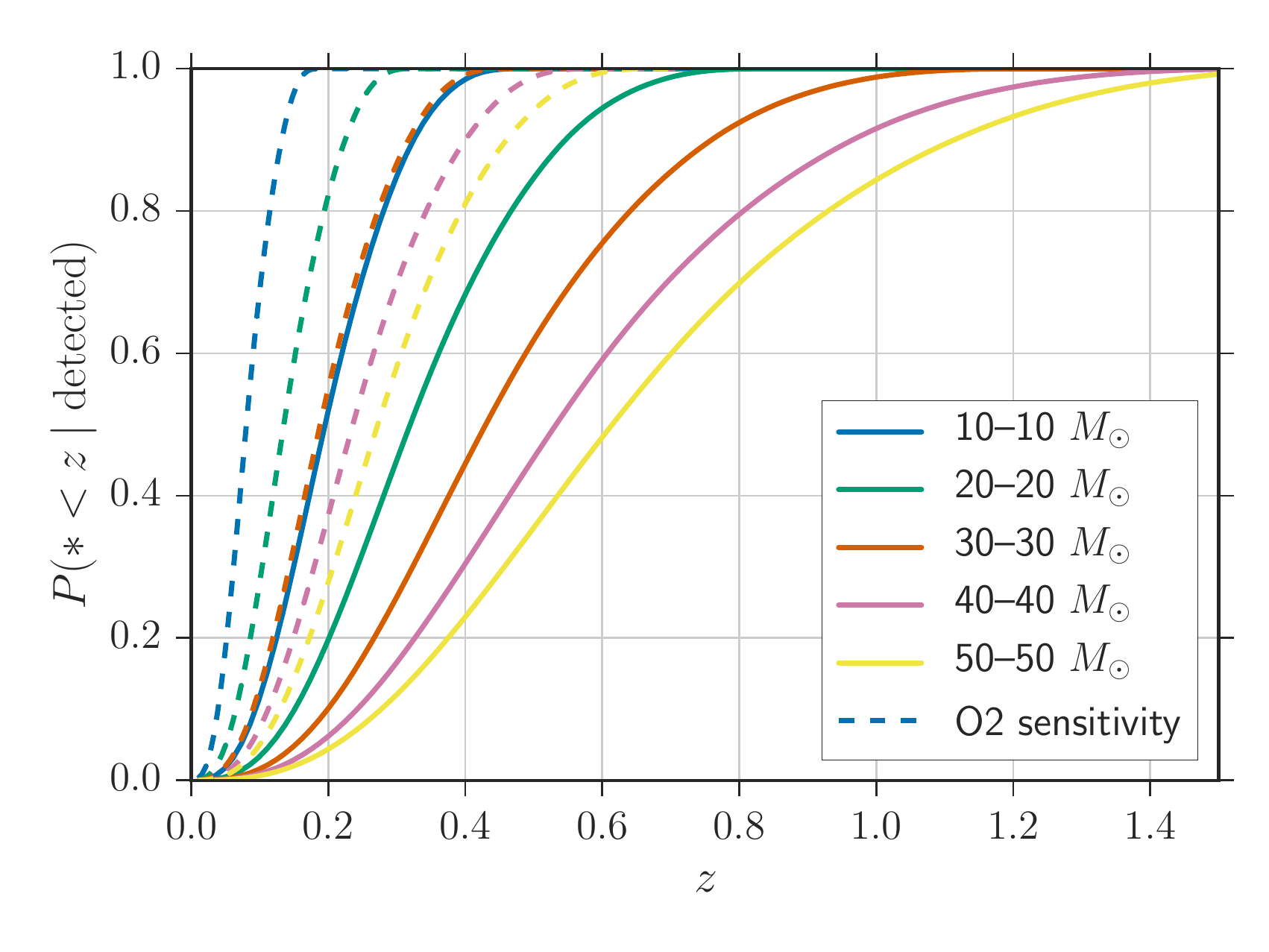}
\caption{Cumulative probability distribution of the redshifts of \emph{detected} BBHs of given masses, assuming that the underlying redshift distribution is uniform in comoving volume. The solid (dashed) lines show the expected distributions for aLIGO at design (O2) sensitivity. If the merger rate evolves positively (negatively) with redshift, these curves would shift to the right (left).}
\end{figure}

The distribution of redshifts among detected BBHs depends on the underlying mass distribution of the black holes. Assuming that the true BBH merger rate is constant in comoving volume, the redshifts of \emph{detected}\/ BBHs follow the cumulative probability distributions shown in Figure~\ref{fig:cumpzdet}, depending on their component masses. 
We assume detected BBHs are those that produce a signal-to-noise ratio (SNR) $\rho > 8$ in a single detector (see \S\ref{sec:fits} for more details).
Within the relevant mass range for stellar-mass BBHs, the GW signal from more massive BBH mergers is intrinsically stronger (``louder'') and can be detected at greater distances. This means that the average redshift among detected heavy BBHs is higher than the average redshift among detected light BBHs. This can be seen in Figure~\ref{fig:cumpzdet}, as the cumulative probability curves shift to the right with increasing BBH mass. Equivalently, the mass distribution of {\em detected}\/ black holes is different from the true underlying mass distribution, with a preference for more massive black holes over less massive ones. Furthermore, as the sensitivity of the GW detector improves, the average redshift of the detected BBHs will increase. This is seen in the difference between the dashed curves, which assume a noise level appropriate to advanced LIGO's (aLIGO's) second observing run, and the solid curves, which assume the noise level for aLIGO at design sensitivity \citep[respectively, the ``Early High Sensitivity" scenario and ``Design Sensitivity'' noise curves from][]{Abbott:OS}. 

It is clear from Figure~\ref{fig:cumpzdet} that it is impossible to infer the underlying BBH redshift distribution independently of the BBH mass distribution. If the BBH merger rate density (rate per comoving volume) increases with increasing redshift, all of the cumulative probability curves in Figure~\ref{fig:cumpzdet} would shift to the right. However, increasing the relative number of massive BBH mergers in the population also increases the proportion of sources detected at high redshift. In other words, the measured redshift distribution alone cannot distinguish between a merger rate that increases with redshift and a population with a high fraction of massive BBHs \citep[see also][]{2018arXiv180204909B}. Likewise, the detected mass distribution is sensitive to the underlying redshift distribution; a greater fraction of detected BBHs will be high mass if the merger rate density increases with redshift, because only the high-mass BBHs are detectable at high redshift. In the following, we jointly examine the mass and redshift distribution of merging BBHs.
\section{Joint Mass-redshift Model}
\label{sec:model}
We consider the differential mass-redshift distribution of BBHs:
\begin{equation}
\label{eq:dN}
\frac{dN}{dm_1dm_2dz} \equiv R\,p(m_1,m_2,z),
\end{equation}
where $R$ is the total number of BBHs across all masses and redshifts, so that:
\begin{equation}
\label{eq:R}
\int \frac{dN}{dm_1dm_2dz} dm_1dm_2dz = R,
\end{equation}
and $p(m_1,m_2,z)$ integrates to unity. 

Given $\frac{dN}{dm_1dm_2dz}$, we can solve for the usual merger rate density, $\frac{dN}{dV_cdt_m}$, where $t_m$ is the source-frame time. The merger rate density as a function of redshift is given by:
\begin{equation}
\label{eq:rate}
\frac{dN}{dV_cdt_m}\left(z\right) = \frac{dN}{dz}\left(z\right) \left[\frac{dV_c}{dz}\left(z\right)\right]^{-1}\frac{1+z}{T_\mathrm{obs}},
\end{equation}
where:
\begin{equation}
\frac{dN}{dz} = \int \frac{dN}{dm_1dm_2dz} dm_1dm_2,
\end{equation}
$T_\mathrm{obs}$ is the total observing time of the GW detector network as measured in the detector frame, and the $(1+z)$ factor converts detector-frame time to source-frame time. 

As a first step, we assume that the underlying mass distribution does not vary across cosmic time, so that we can factor the joint mass-redshift distribution as:
\begin{equation}
\label{eq:pmz}
p(m_1,m_2,z) = p(m_1,m_2)p(z).
\end{equation}
This assumption may break down over a large range of redshifts, as many formation scenarios predict some dependence of the mass distribution on the merger redshift. However, aLIGO is only sensitive to redshifts $z \lesssim 1.5$ (see Figure~\ref{fig:cumpzdet}), where Equation~\ref{eq:pmz} is likely a good approximation, particularly if the distribution of delay times is broad \citep[see, for example, Figure~3 in][]{Mapelli:2017,2041-8205-836-2-L26}. 
For the mass distribution, we use the two-parameter model from \cite{2017ApJ...851L..25F}, which is an extension to the power-law model employed by the LVC to fit the BBH mass distribution to the first four detections \citep{2016PhRvX...6d1015A,2017PhRvL.118v1101A} incorporating the possibility of a mass gap above $\gtrsim 40\,M_\odot$ due to pair-instability supernovae~\citep{1964ApJS....9..201F,2002ApJ...567..532H,2016A&A...594A..97B}.
We assume that the mass distribution takes the form:
\begin{equation}
\label{eq:pm}
p(m_1, m_2 \mid \alpha, M_\mathrm{max}) \propto \frac{m_1^{-\alpha}}{m_1-5\ M_\odot}\mathcal{H}(M_\mathrm{max} - m_1),
\end{equation}
where $\mathcal{H}$ is the Heaviside step function. We fix the distribution of secondary masses, $m_2$, to be uniform between the minimum BH mass and $m_1$, and fix the minimum BH mass, $M_\mathrm{min} = 5 \ M_\odot$.
\subsection{Redshift Model A}
For our first redshift model, we choose the following parametrization:
\begin{equation}
\label{eq:uc_dev}
p(z \mid \lambda) \propto \frac{dV_c}{dz}\frac{1}{1+z}(1+z)^\lambda,
\end{equation}
so that $\lambda = 0$ reduces to a merger rate density that is uniform in comoving volume and source-frame time. The extra factor of $(1+z)^{-1}$ converts from detector-frame to source-frame time. \maya{Note that for very small $z$, Equation~\ref{eq:uc_dev} reduces to a constant in comoving volume and source-frame time merger rate regardless of the value of $\lambda$.}

If the rate density follows the specific SFR, we would expect:
\begin{equation}
\label{eq:sfr_rate}
p(z) \propto \frac{dV_c}{dz}\frac{1}{1+z}\psi(z),
\end{equation}
where $\psi(z)$ is the specific SFR \citep{MD:2014}:
\begin{equation}
\psi(z) = 0.015 \frac{(1+z)^{2.7}}{1+\left[(1+z)/2.9\right]^{5.6}} \,{M}_\odot\,\rm{yr}^{-1}\,\rm{Mpc}^{-3}.
\end{equation}
Other models for the SFR, such as \cite{doi:10.1093/mnras/stu2600} or \cite{2005JCAP...04..017S}, agree with the Madau-Dickinson SFR at the low redshifts relevant to aLIGO, $z < 1.5$. We note that Equation~\ref{eq:uc_dev} with $\lambda = 2.7$ approximates Equation~\ref{eq:sfr_rate} for $z \ll 1$, whereas $\lambda = 2.4$ provides a very good approximation to Equation~\ref{eq:sfr_rate} for $0.1 \lesssim z\lesssim 1$.
Alternatively, because BBH formation is more efficient at low metallicities, we might expect that the rate density follows the low-metallicity SFR, $\psi(z)f_Z(z)$, where $f_Z(z)$ is the fraction of star formation occurring at metallicity $\leq Z$ at redshift $z$ \citep{Belczynski:2010,2013MNRAS.429.2298M,2015MNRAS.451.4086S,2016ApJ...818L..22A,2016MNRAS.458.2634M}. For example, \cite{2006ApJ...638L..63L} give the fit:
\begin{equation}
\label{eq:CDF_Zz}
f_Z(z) = \hat{\Gamma}\left( 0.84, \frac{Z}{Z_\odot}^2 10^{0.3z}\right),
\end{equation}
where $\hat{\Gamma}$ is the incomplete gamma function \citep[see also][]{2016MNRAS.458.2634M}. As the average metallicity decreases with increasing redshift, the low-metallicity SFR rises more steeply with increasing redshift, and peaks at higher redshift. We find that a rate density that follows the low-metallicity ($Z \leq 0.3Z_\odot$) SFR:
\begin{equation}
\label{eq:lowZSFR}
p(z) \propto \frac{dV_c}{dz}\frac{1}{1+z}\psi(z)f_{Z=0.3Z_\odot}(z),
\end{equation}
leads to a redshift distribution, $p(z)$, that is well approximated by Equation~\ref{eq:uc_dev} with $\lambda = 3.3$. It has also been proposed that the progenitors of BBHs are Population III stars formed at zero metallicity, in which case we might expect an even steeper increase of the merger rate with increasing redshift \citep{Belczynski:2004,Kinugawa:2014}.

More realistically, the rate density follows the SFR convolved with a time-delay distribution. Different formation channels predict different time-delay distributions. If typical time delays are very long ($\sim$4--11 Gyr), as in the chemically homogeneous formation channel, the rate density will peak at very low redshifts ($z \sim 0.4$) well within the aLIGO horizon \citep{2016MNRAS.458.2634M}. Under the parametrization of Equation~\ref{eq:uc_dev}, this corresponds to $\lambda < 0$; in the range $z\leq 1$, the best fits to such redshift distributions are given by $-6 \leq \lambda \leq -4$. In the classical field formation scenario, typical time delays are much shorter ($\sim$10--300 Myr) \citep{2012ApJ...759...52D,2013ApJ...779...72D}. In this field formation channel, the time delay is expected to follow a distribution:
\begin{equation}
\label{eq:td}
\begin{aligned}
\tau \propto \tau^{-1} && \tau_\mathrm{min} < \tau < \tau_\mathrm{max},
\end{aligned}
\end{equation}
where typically $\tau_\mathrm{min} \sim 50$ Myr and $\tau_\mathrm{max}$ is a Hubble time. In the redshift range of interest to aLIGO, this corresponds to a merger density that increases with increasing redshift $(\lambda > 0)$, but is less steep than the SFR. \maya{For example, if the formation rate of BBHs follows the Madau-Dickinson SFR, and the time-delay between formation and merger follows Equation~\ref{eq:td}, the merger rate at $z \lesssim 1$ can be described by Equation~\ref{eq:uc_dev} with $\lambda \sim 1.3$. A measurement of $\lambda > 1.3$ for the merger rate would indicate that the BBH formation rate density peaks at higher redshift than the SFR (possibly because of metallicity evolution), or that there is a stronger preference for very short time delays.}

If the time-delay distribution is in fact restricted to very short time delays (for example, a flat distribution between $\tau_\mathrm{min}=50$ Myr and $\tau_\mathrm{max} = 1$ Gyr), the BBH merger rate density will be nearly identical to their formation rate density at $z \lesssim 1$. This may be the case for mergers that take place inside globular clusters. On the other hand, binaries that form dynamically inside clusters but are ejected prior to merger tend to have much longer time delays, on the order of $\sim 10$ Gyr \citep{2016PhRvD..93h4029R}, corresponding to $\lambda \sim -10$.

\subsection{Redshift Model B}
\begin{figure}
\label{fig:pzdet}
\includegraphics[width=0.5\textwidth]{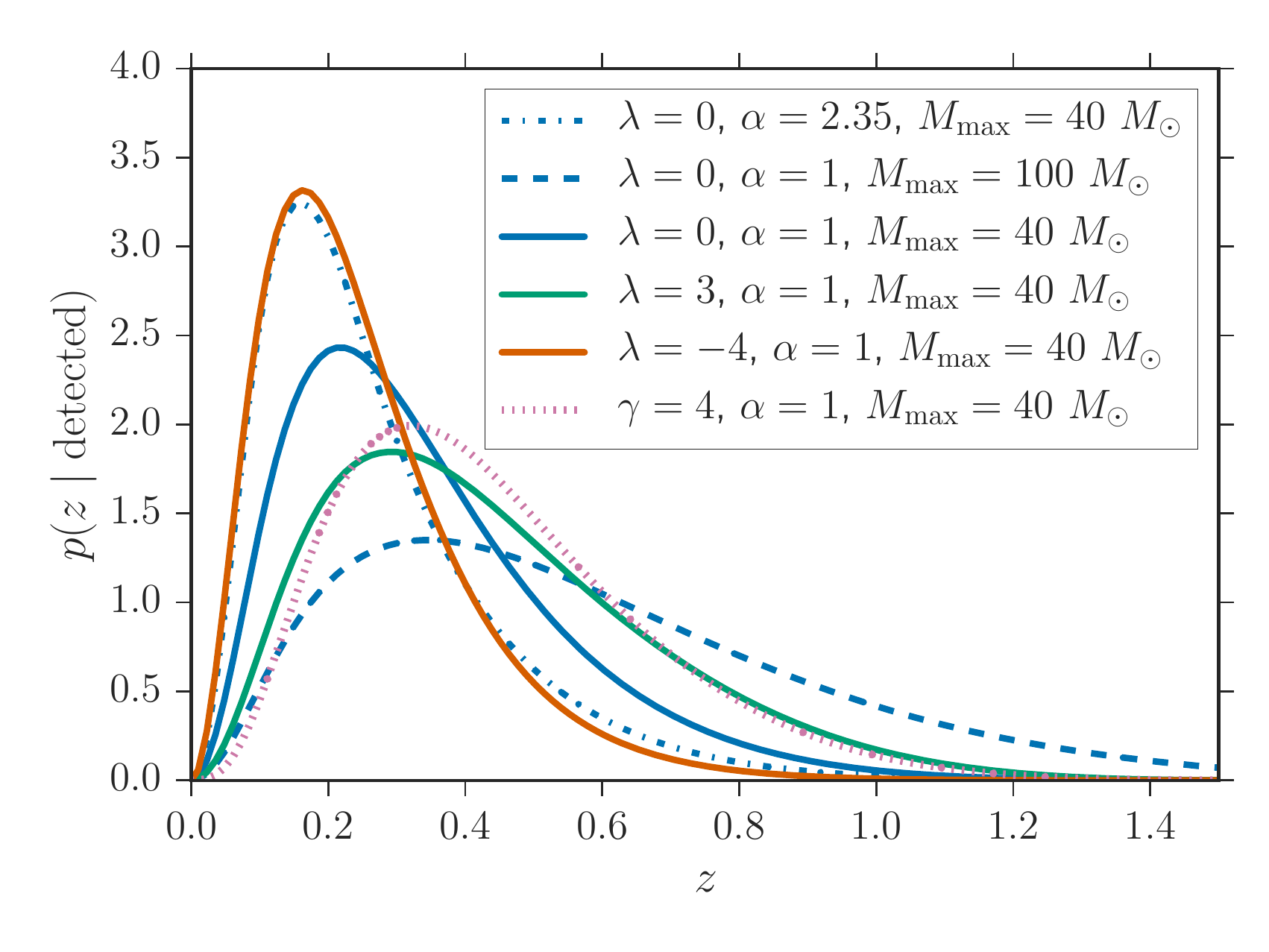}
\caption{Expected redshift distributions among the \emph{detected} BBHs for LIGO-Virgo operating at design sensitivity, for different choices of the underlying redshift distribution parametrized by $\lambda$ (Model A) or $\gamma$ (Model B). The detected redshift distributions depend on the underlying mass distribution, which we parametrize with a power-law slope, $\alpha$, and an upper mass cutoff, $M_\mathrm{max}$.}
\end{figure}

For our second redshift model, we assume that the merger rate distribution is uniform in $V_c^{\gamma/3} = D_c^{\gamma}$ and source-frame time, so that $\gamma = 3$ implies that the distribution is uniform in comoving volume. In other words, the redshift distribution takes the form:
\begin{equation}
p(z \mid \gamma) \propto \frac{1}{1+z}\frac{D_c^{\gamma-1}}{E(z)}, 
\end{equation}
where $D_c$ is the comoving distance and $dD_c \propto \frac{dz}{E(z)}$ \citep{Hogg:2000}. 
This redshift model is more flexible than Model A in the local universe, as it allows for large deviations from a constant in volume merger rate at low redshifts, whereas Model A always reduces to a constant in volume merger rate in the limit $z \to 0$.
Model B can constrain scenarios that would cause extreme variations in the slope of the redshift distribution locally, such as a significant population of strongly lensed sources that appear closer than they are \citep{2018arXiv180205273B}, leading us to infer $\gamma < 3$, or GW leakage causing sources to appear farther than they are, leading us to infer $\gamma > 3$ \citep{2007ApJ...668L.143D}. Previous studies have explored such scenarios through their effects on the SNR distribution \citep{Chen:2014,2016JCAP...07..021G,2016CQGra..33p5004C} or GW standard siren measurements \citep{2018arXiv180108160P}.
However, such effects on the redshift distribution (and likewise, the SNR distribution) will most likely be difficult to disentangle from the astrophysical processes that control the redshift and mass distributions. For example, in Figure~\ref{fig:pzdet}, the dotted pink curve, corresponding to Redshift Model B, $\gamma = 4$, and the solid green curve, corresponding to Redshift Model A, $\lambda = 3$, are very similar.

Figure~\ref{fig:pzdet} shows the expected redshift probability density function (PDF) for sources detected by aLIGO at design sensitivity, assuming that the true mass-redshift distribution is described by the models discussed in this section. The solid, dashed, and dashed-dotted blue curves assume the same underlying redshift distribution (corresponding to a constant merger rate density), but different mass distributions. Meanwhile, the solid blue, orange, and green curves show how the detected redshift PDF varies with different underlying redshift distributions parametrized by $\lambda$, for a fixed mass distribution. The dotted pink curve assumes the same mass distribution, but takes the underlying redshift distribution to follow Model B with $\gamma = 4$. If the merger rate increases with redshift (for example, the solid green compared to the solid blue curve), the detected distribution will skew to high redshifts. However, the effects of changing the mass distribution can be equally, if not more, significant (for example, the difference between the dashed, dashed-dotted and solid blue curves). As we shall see in the following section when we infer the parameters of the mass-redshift model, this leads to a degeneracy between the mass parameters and the redshift evolution parameter.

\section{Fitting the Mass-redshift Distribution}
\label{sec:fits}
In this section, we fit our parametrized model for the differential mass-redshift distribution, $\frac{dN}{dm_1dm_2dz}$, to real and simulated LIGO-Virgo detections. Our goal is to extract the four population parameters of the model from GW measurements of the masses and luminosity distances of detected sources. We assume a fixed $\Lambda$CDM cosmology determined by the 2015 Planck cosmological parameters~\citep{PlanckCosmology}, so that the measured luminosity distance is a direct measurement of the redshift. The shape of the mass-redshift distribution is governed by three parameters, $\boldsymbol{\theta}$. For Model A of the redshift evolution, $\boldsymbol{\theta} = \lbrace \alpha, M_\mathrm{max}, \lambda \rbrace$ and for Model B, $\boldsymbol{\theta} = \lbrace \alpha, M_\mathrm{max}, \gamma \rbrace$. The fourth parameter, $R$, corresponds to the total number of detected BBH systems and gives the overall normalization according to Equation~\ref{eq:R}, allowing us to solve for the physical merger rate of BBHs (Equation~\ref{eq:rate}).

We model the rate density $\frac{dN}{dm_1dm_2dz}$ as a Poisson point process \citep{Loredo:2004,2014arXiv1412.4849F,Wysocki:2018}. The likelihood for the GW data $\lbrace d_i \rbrace_{i=1}^{N_\mathrm{obs}}$ from $N_\mathrm{obs}$ observations, given population parameters $\lbrace \boldsymbol{\theta}, R \rbrace$ is given by:
\begin{widetext}
\begin{align}
\label{eq:likelihood}
p(\lbrace d_i \rbrace \mid \boldsymbol{\theta}, R) &= \left[ \prod_{i=1}^{N_\mathrm{obs}}  \int p(d_i \mid m_1,m_2,z) \frac{dN}{dm_1dm_2dz}\left(m_1,m_2,z \mid \boldsymbol{\theta},R\right) dm_1dm_2dz \right] e^{-\beta(\boldsymbol{\theta},R)}\\
&= \left[ \prod_{i=1}^{N_\mathrm{obs}} \left\langle \frac{1}{\pi(z^i,m_1^i,m_2^i)}\frac{dN}{dm_1dm_2dz}\left(m_1^i,m_2^i,z^i \mid \boldsymbol{\theta},R\right)  \right\rangle_{\lbrace z^i, m_1^i, m_2^i \rbrace } \right] e^{-\beta(\boldsymbol{\theta},R)},
\end{align}
\end{widetext}
where $\langle \cdots \rangle$ denotes an average over $\lbrace z^i, m_1^i, m_2^i \rbrace$ posterior samples from the $i$th event and $\pi(z^i, m_1^i, m_2^i)$ denotes the interim prior used in the analysis of individual events. The standard priors used in the LIGO analysis of individual events are uniform in component masses and ``volumetric'' in distance \citep{Veitch:2015}:
\begin{equation}
\label{eq:Euclideanprior}
\pi(z,m_1,m_2) \propto d_L(z)^2.
\end{equation}
\maya{It is interesting to note that the redshift distribution described by Equation \ref{eq:Euclideanprior} matches Model A with $\lambda = 3$ (as opposed to $\lambda = 0$, which corresponds to a constant rate density), and so the simplifying assumption that the universe follows a Euclidean geometry implies a redshift distribution that mimics the SFR.} Meanwhile, $\beta (\boldsymbol{\theta},R)$ is given by:
\begin{equation}
\label{eq:beta}
\beta (\boldsymbol{\theta},R) = \int \frac{dN}{dm_1dm_2dz}\left(m_1,m_2,z \mid \boldsymbol{\theta},R\right)P_\mathrm{det}(m_1,m_2,z)dm_1dm_2dz
\end{equation}
where $P_\mathrm{det}(z, m_1,m_2)$ is the fraction of binary sources at a given redshift and of given component masses that are detectable by the GW detector network. We assume that sources are isotropically distributed on the sky, and the binary inclination is uniformly distributed on the sphere (i.e., uniform in $\cos(\mbox{inclination})$). We also fix all BH spins to zero in our analysis. Alternatively, we could allow the spins to vary and marginalize over the spin distribution when measuring the mass-redshift distribution \citep{Wysocki:2018}. However, incorporating the spin distribution will not affect our analysis significantly, considering that BBHs seem to have small aligned-spin components \citep{2017Natur.548..426F, 2018ApJ...854L...9F}. 

Given the component masses and spins, sky position, inclination, and distance (or equivalently, redshift) of the BBH source relative to a GW detector, together with a power spectral density (PSD) that characterizes the noise of the detector, we can calculate the single-detector SNR \citep{FinnChernoff:1993,2015ApJ...806..263D}. We consider a single-detector SNR threshold $\rho_\mathrm{th} = 8$ for detection (corresponding to a network SNR threshold of 12), and assume that for O1 and O2, the noise follows the PSD given by the aLIGO ``Early High Sensitivity" scenario \citep{Abbott:OS}. For this calculation we ignore the distinction between the true and measured SNR, which, with only six events, does not significantly impact our results. The probability of detection, $P_\mathrm{det}(z, m_1,m_2)$, is therefore the fraction of sources that produce a true SNR of $\rho = 8$ in a single detector.

We are interested in the posterior probability of the population parameters $\lbrace \boldsymbol{\theta}, R \rbrace$, which is related to the likelihood in Equation~\ref{eq:likelihood} by a prior:
\begin{equation}
\label{eq:posterior}
p(\boldsymbol{\theta}, R \mid \lbrace d_i \rbrace) \propto p(\lbrace d_i \rbrace \mid \boldsymbol{\theta}, R)p(\boldsymbol{\theta}, R).
\end{equation}
We choose broad, uninformative priors. We take a flat prior for the parameters that make up $\boldsymbol{\theta}$ (the power-law slope, maximum component mass, and redshift evolution parameter). Our default prior ranges are $\alpha \in [-4,5]$, $M_\mathrm{max} \in [31 \ M_\odot, 100 \ M_\odot]$, $\lambda \in [-50,30]$ and $\gamma \in [0,8]$. For the rate parameter $R$, we take a flat-in-log prior over the range $R \in [10,10^{12}]$. Recall that $R$ is the total number of mergers between redshift $z=0$ and the maximum redshift at which the detectors are sensitive---roughly $z=0.6$ for O1 and O2 \citep{2016ApJ...818L..22A,Kissel:2016}. The combined prior is then:
\begin{equation}
p(\boldsymbol{\theta}, R) \propto \frac{1}{R}.
\end{equation}
With this prior choice, the posterior marginalized over $R$ reduces to:
\begin{equation}
\label{eq:postnoR}
p(\boldsymbol{\theta} \mid \lbrace d_i \rbrace) \propto \prod_{i=1}^{N_\mathrm{obs}} \frac{\int p(d_i \mid m_1,m_2,z) p\left(m_1,m_2,z \mid \boldsymbol{\theta}\right) dm_1dm_2dz}{\xi\left(\boldsymbol{\theta}\right)} p(\boldsymbol{\theta}),
\end{equation}
where:
\begin{align}
\label{eq:xi}
\xi\left(\boldsymbol{\theta}\right) &= \int p\left(m_1,m_2,z \mid \boldsymbol{\theta}\right)P_\mathrm{det}(m_1,m_2,z)dm_1dm_2dz \\
&= {\beta(\boldsymbol{\theta},R)}/{R},
\end{align}
and $p\left(m_1,m_2,z \mid \boldsymbol{\theta}\right)$ is related to $\frac{dN}{dm_1dm_2dz}(m_1,m_2,z, \mid \boldsymbol{\theta},R)$ by Equation~\ref{eq:dN}. Equation~\ref{eq:postnoR} follows because Equation~\ref{eq:posterior} can be written as:
\begin{equation}
p(\boldsymbol{\theta}, R \mid \lbrace d_i \rbrace) \propto p\left(\boldsymbol{\theta} \mid \lbrace d_i \rbrace\right)\left[\xi\left(\boldsymbol{\theta}\right)\right]^{N_\mathrm{obs}}R^{N_\mathrm{obs}-1}e^{-R\xi\left(\boldsymbol{\theta}\right)},
\end{equation}
which when marginalized over $R$, yields $(N_\mathrm{obs}-1)!\, p\left(\boldsymbol{\theta} \mid \lbrace d_i \rbrace\right)$. Equation~\ref{eq:postnoR} is identical to the form of the posterior derived in \cite{Mandel:2016} and used in previous population analyses of GW events \citep{Abbott:rates,2017ApJ...851L..25F}.

\subsection{LIGO-Virgo detections}
\label{sec:detections}

\begin{figure*}
\label{fig:lambda-alpha_post_mock6}
\includegraphics[width=\textwidth]{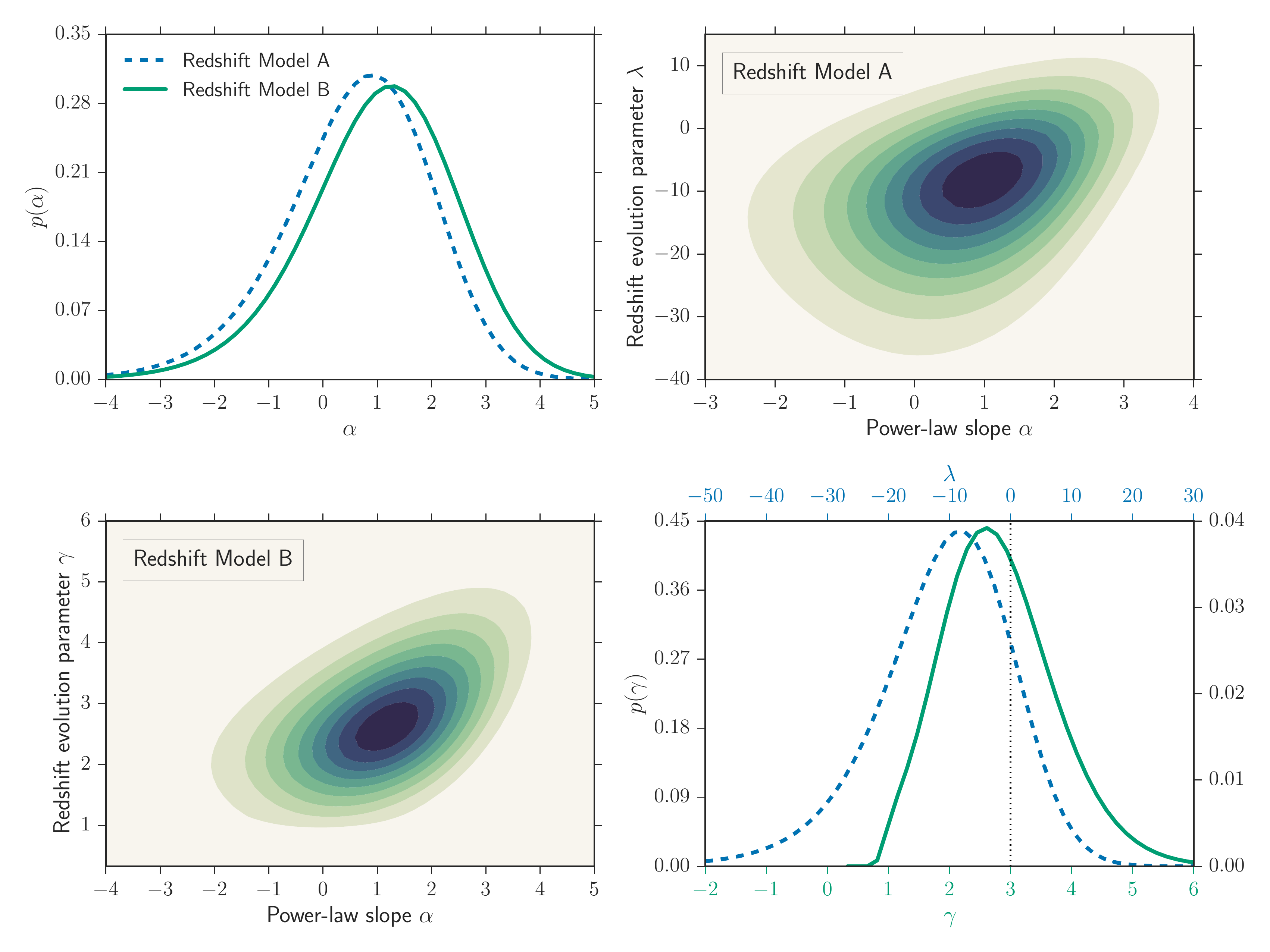}
\caption{Posterior PDF of the the power-law slope, $\alpha$, and the redshift evolution parameter from Model A ($\lambda$) and Model B ($\gamma$)  from the first six announced BBH detections. The top right (bottom left) panel shows the two-dimensional posterior on $\alpha$ and $\lambda$ ($\gamma$), calculated from the full posterior $p(\boldsymbol{\theta},R \mid \mathbf{d})$ marginalized over $M_\mathrm{max}$ and $R$. The contours show increasing probability in 10\% steps. The top left panel shows the posterior on $\alpha$ marginalized over all other parameters, for both Model A (dashed blue curve) and Model B (solid green curve) of the redshift evolution. The bottom right panel shows the posterior PDF for the redshift evolution parameters for the two models. The first six announced LIGO-Virgo detections are consistent with a uniform rate density ($\lambda = 0$ or $\gamma = 3$; dotted black line in bottom right panel) within the 68\% credible interval, at the 56\% (34\%) credible level enclosing the maximum a posteriori value for Model A (Model B).}
\end{figure*}

\begin{figure*}
\label{fig:dNdz}
\includegraphics[width=\textwidth]{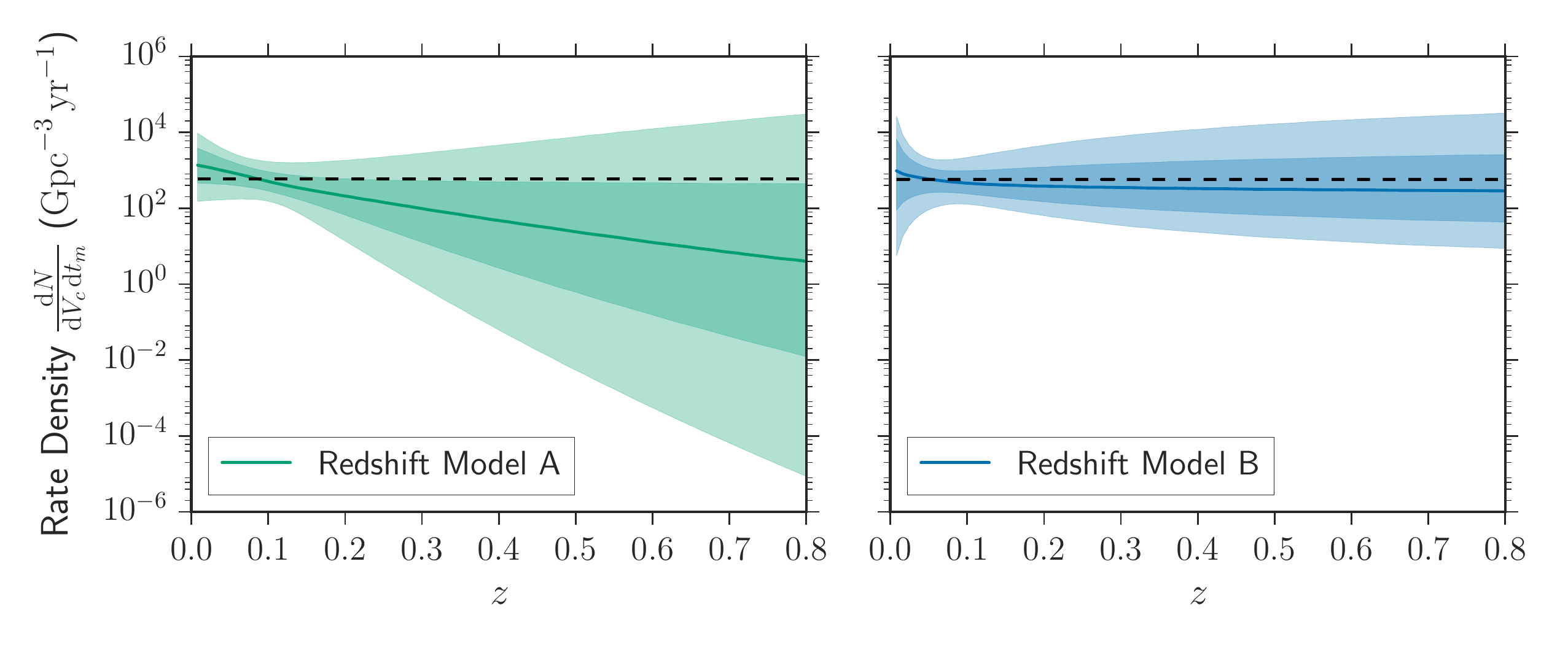}
\caption{Merger rate density as a function of redshift for Model A (left) and Model B (right) of the redshift evolution, assuming that the six published LIGO-Virgo detections form a complete sample, and were detected during a 94-day observing period. The solid line shows the median rate density as a function of redshift, and the light and dark shaded regions show equal-tail 68\% and 95\% credible levels, respectively. Our inferred merger rate is consistent (at the 68\% credible level) with being uniform in comoving volume and source frame time, $t_m$, which corresponds to a flat horizontal line on this plot (dashed black line). \maya{Our analysis shows a preference for a merger rate density that decreases with increasing redshift; however, this may be due to a false assumption that the six published BBHs form a complete sample from O1 and O2, as discussed in the text. Proper analysis, using the final sample and correct observing time, should wait for the analysis of O2 data to officially conclude.}
}
\end{figure*}

We fit our mass-redshift model to the first six announced BBH detections. \maya{We caution that at the time of writing, the analysis of LIGO's second observing run is still ongoing, and the sample of detections is not guaranteed to be complete. In order to avoid introducing any unmodeled selection biases, proper analysis should wait until the release of the final sample; nevertheless, our analysis illustrates the types of constraints we expect from six BBHs.}

From each of these six events, we approximate the mass and redshift posterior PDFs using the published central values and 90\% credible bounds.
Specifically, we approximate the \emph{detector}-frame chirp mass posterior PDFs by Gaussian distributions with a mean and standard deviation that match the published medians and 90\% credible widths \citep{2016PhRvX...6d1015A,2017PhRvL.118v1101A,PhysRevLett.119.141101,2017ApJ...851L..35A}. The detector-frame masses differ from the source-frame masses by a factor of $(1+z)$ \citep{1987GReGr..19.1163K,2005ApJ...629...15H}. 
Similarly, we use the published medians and 90\% credible bounds on the mass ratio, $q = \frac{m_2}{m_1}$, to find the median and 90\% credible bounds on the symmetric mass ratio:
\begin{equation}
\eta \equiv \frac{q}{(1+q)^2}.
\end{equation}
We then approximate the symmetric mass-ratio posteriors by Gaussian distributions with means and standard deviations that match these medians and 90\% credible intervals. We use these approximate posteriors on chirp mass and symmetric mass ratio to approximate the detector-frame component mass posterior distributions for each event. Lastly, we approximate the redshift posterior for each event by a Gaussian distribution matching the published median and 90\% credible intervals. We therefore generate posterior samples for the detector-frame masses and redshifts following these approximate distributions. To get posterior samples for the \emph{source}-frame masses, $m_1$ and $m_2$, we divide the posterior samples for the detector-frame masses by $(1+z)$, where the redshifts, $z$, are drawn from the redshift posterior distribution. This captures the correlations between redshift and source-frame masses in the posterior PDF for an individual event, $p(m_1, m_2, z \mid d_i)$.

Figure~\ref{fig:lambda-alpha_post_mock6} shows the resulting posterior PDF on the power-law slope, $\alpha$, and the redshift evolution parameter, $\lambda$ or $\gamma$, marginalized over $M_\mathrm{max}$ and $R$:
\begin{equation}
p(\lambda, \alpha \mid \lbrace d_i \rbrace) = \int p(\lambda, \alpha, M_\mathrm{max}, R \mid \lbrace d_i \rbrace)dM_\mathrm{max}dR.
\end{equation}

Under Model A, we infer $\lambda = -10^{+15}_{-21}$, $\alpha = 0.7 \pm 2.0$, and under Model B, we infer $\gamma = 2.7^{+1.8}_{-1.3}$, $\alpha = 1.1^{+2.1}_{-2.5}$. All credible intervals are quoted as the median and symmetric (equal-tailed) 90\% range.
Meanwhile, from the marginal posterior PDF on $M_\mathrm{max}$, we infer that the maximum component BH mass is $39^{+30}_{-6} \ M_\odot$ (Model A) or $39^{+28}_{-6} \ M_\odot$ (Model B); the 95\% upper limit of $\sim69 \ M_\odot$ is tighter than the 95\% upper limit of $\sim77 \ M_\odot$ found in~\cite{2017ApJ...851L..25F} with the additional two detections analyzed here.

There is a positive correlation between the mass power-law slope and the redshift evolution parameter in the two-dimensional posterior, because a mass distribution that favors low masses (large $\alpha$) is compatible with the data only if the merger rate increases with redshift (large $\lambda$ or $\gamma$) and vice versa; otherwise, more high-redshift and more low-mass objects would have been detected. The inferred rate parameter, $R$, is also positively correlated with these parameters, large $\alpha$ and/or $\lambda$ (equivalently, $\gamma$) imply that there are many more high-redshift low-mass sources that contribute to the total number of mergers $R$, but not to the detected number, $N_\mathrm{obs}$ \citep[see also][]{Wysocki:2018}.

\maya{It has recently been suggested that there are statistically too many nearby BBH detections (or equivalently, too many high-SNR detections) compared to the expected constant in comoving volume distribution \citep{2018arXiv180205273B}. Although our analysis shows a slight preference for a merger rate density that declines with increasing redshift, we find that for both models A and B of the redshift evolution, the current data is consistent with a uniform in comoving volume rate density ($\lambda = 0$ or $\gamma = 3$) within $1\sigma$ (68\% credibility). Furthermore, it is possible that the set of published LIGO-Virgo detections is incomplete at this time. If loud events are published first, it is possible that an incomplete set would be biased toward low-redshift events, and our analysis would be artificially biased to small values of $\lambda$ and $\gamma$. A more complete analysis can take place once the results from LIGO's second observing run are finalized.}

Finally, we can calculate a posterior PDF on the merger rate density as a function of redshift according to Equation~\ref{eq:rate}. As an illustration of the method, we assume that the total observing time from aLIGO's first and second observing runs is 94 days. This assumption is not based on the true observing time, which is not yet known as analysis on O2 is ongoing. Instead, it is chosen to match the most recently published merger rate estimate from LIGO in~\cite{2017PhRvL.118v1101A}, which is in the range [12, 213] Gpc$^{-3}$ yr$^{-1}$ for a BBH population with a uniform in comoving volume merger rate and a mass distribution with power-law slope $1 \leq \alpha \leq 2.35$. With six published detections, an observing time of 94 days yields a mean merger rate of 100 Gpc$^{-3}$ yr$^{-1}$ for the ``power-law'' ($\alpha = 2.35$) population considered in~\cite{2017PhRvL.118v1101A}. (Note that this ``power-law'' mass distribution from~\cite{2017PhRvL.118v1101A} fixes the maximum component BH mass to $M_\mathrm{max} = 95 \ M_\odot$, the minimum component BH mass to $M_\mathrm{min} = 5 \ M_\odot$, and the maximum total binary mass to $M_\mathrm{tot, max} = 100 \ M_\odot$.) With this assumption of the observing time, Figure~\ref{fig:dNdz} shows the inferred rate density (marginalized over all population parameters) as a function of redshift for Model A (left panel) and Model B (right panel) of the redshift evolution. While previously published merger rate estimates are valid only for a fixed redshift distribution, this method allows us to infer the merger rate simultaneously with the mass and redshift distributions. Once again, we see that our results are consistent with a non-evolving merger rate, which would correspond to a flat horizontal line in Figure~\ref{fig:dNdz}.

From Figure~\ref{fig:dNdz}, we see that the merger rate is well-constrained at redshifts $0.05 \lesssim z \lesssim 0.15$. Meanwhile, there is too little volume at $z \lesssim 0.05$ to constrain the merger rate well, and the detectors are not sensitive enough at high redshifts. The uncertainties on the merger rate become especially large at high redshifts for Model A, because the parameter for this model, $\lambda$, is hard to constrain with low-redshift observations (where the merger rate always approaches a constant in comoving volume rate) but has a significant effect on the high-redshift rate. Meanwhile, varying the parameter, $\gamma$, in Model B causes a large variation in the low-redshift rate. This means that $\gamma$ is easier to measure with low-redshift observations than $\lambda$, and so assuming Model B, the merger rate is relatively well constrained at high redshifts where the sensitivity of the GW detectors approaches zero. 

\subsection{Future detections}
\label{sec:future}
In this section, we simulate detections from a mock population of BBHs and apply our method to infer the underlying mass and redshift distribution parameters. Our goal is to estimate how many LIGO-Virgo detections will be required to correctly infer a deviation from a uniform in comoving volume merger rate, or alternatively, how many detections will be required to confidently rule out strong deviations from a uniform in comoving volume merger rate. We consider two simulated populations. Both populations follow the same distribution for the BBH masses (Equation~\ref{eq:pm}), with a minimum component mass of $5 \ M_\odot$, a maximum component mass of $40 \ M_\odot$, and a power-law slope $\alpha = 1$. We assume that the mass distribution for both populations is independent of the redshift distribution (Equation~\ref{eq:pmz}). The redshift distribution of the first population follows Model A (Equation~\ref{eq:uc_dev}) with $\lambda = 3$, as might be expected if the merger rate followed the low-metallicity SFR convolved with a time-delay distribution that favored short time delays. Meanwhile, the second population has a uniform in comoving volume merger rate, corresponding to $\lambda = 0$ in Model A or $\gamma = 3$ in Model B. 

\begin{figure*}
\label{fig:simconstraints}
\includegraphics[width=\textwidth]{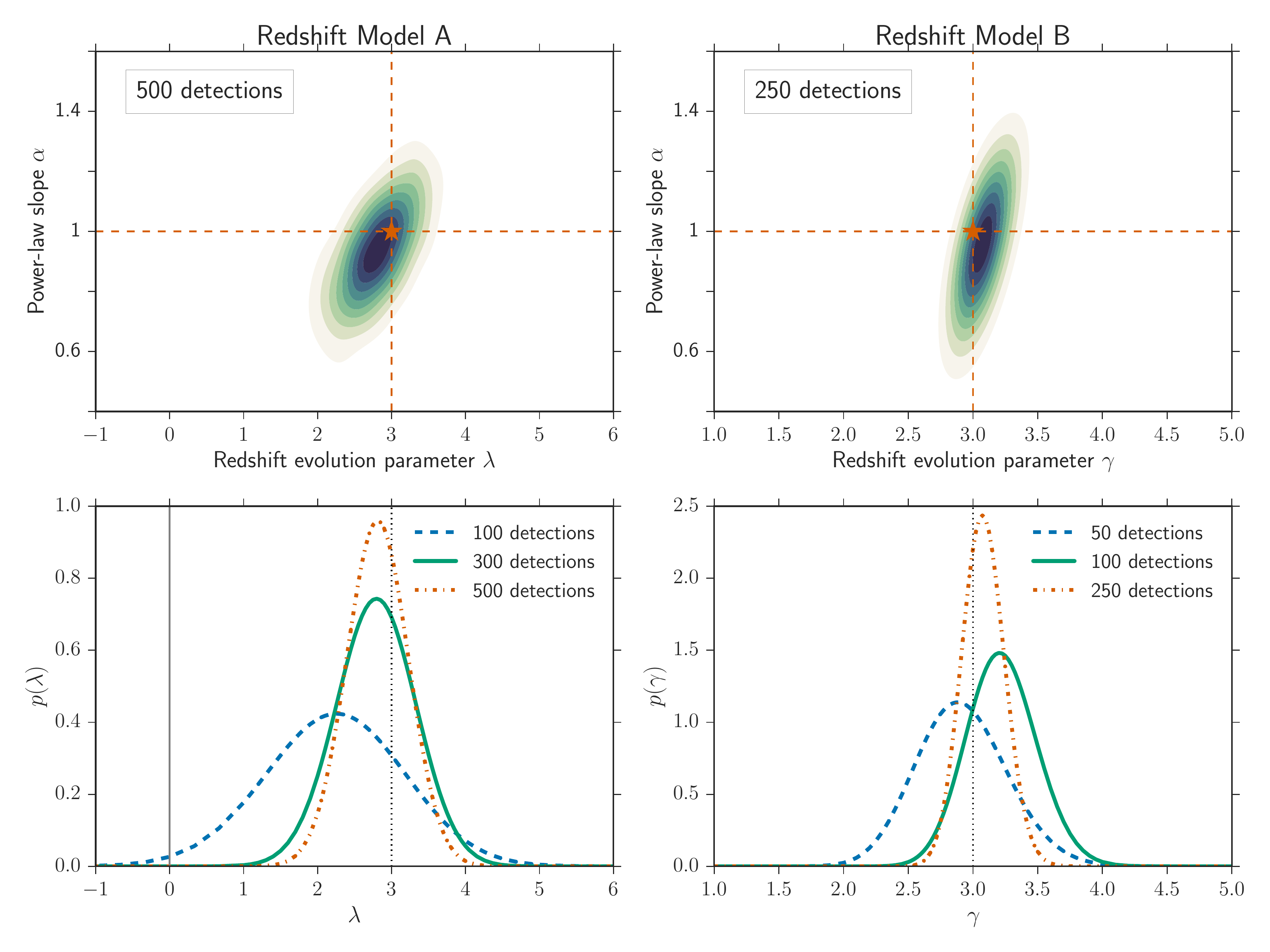}
\caption{Projected constraints on the mass power-law slope and the redshift evolution parameter for a set of simulated BBH detections from two populations. The top panels show the joint posterior PDF on the power-law slope and redshift evolution parameter, marginalized over the rate and maximum mass parameters. The bottom panels show the marginalized posterior on the redshift evolution parameter. Both populations follow the same mass distribution described by $\alpha = 1$, $M_\mathrm{max} = 40 \ M_\odot$, but differ in their redshift distribution. \emph{Left panels:} This population is described by a redshift distribution that roughly follows the SFR, or $\lambda = 3$ in Model A. After 100 detections by LIGO-Virgo at design sensitivity (solid green line, bottom left panel), the constraints on $\lambda$ are tight enough to exclude a uniform in comoving volume merger rate, $\lambda = 0$ (black solid line), at 99\% credibility. \emph{Right panels:} This population has a uniform in comoving volume merger rate ($\gamma = 3$ in Model B). We fit detections from this population with Model B of the redshift evolution, and the parameter $\gamma$ is sufficiently well-constrained after 100 detections to constrain $\gamma$ to a 90\% credible interval of $\lesssim 1$.}
\end{figure*}

For each mock population we generate BBHs with component masses and redshifts following the assigned underlying distribution. Only a subset of BBHs are detected, and their measured masses and redshifts take the form of (marginalized) posterior PDFs. In order to generate realistic mass and redshift measurements, we construct a synthetic detection model. The synthetic detection model enables us to self-consistently and realistically capture the correlations between the measured SNR, which determines the detectability of an event and its measured redshift. We verify that the model closely approximates the detectability of a BBH with given masses and redshift by aLIGO at design sensitivity.

Synthetic BBHs are generated as follows. Each BBH system is characterized by four parameters: the source-frame chirp mass $\mathcal{M}$, the symmetric mass ratio $\eta$, the luminosity distance $d_L$, and an angular factor $\Theta$. Each system also has an associated \emph{true} SNR, $\rho$, which depends on these four parameters. Note that $\mathcal{M}$ and $\eta$ allow us to directly infer $m_1$ and $m_2$, and $d_L$ allows us to infer $z$. Here, $\Theta$ plays the combined role of the sky location, inclination, and polarization on the measured GW amplitude. We tune the width of the $\Theta$ distribution to control the uncertainty of the \emph{measured} signal strength, which in turn controls the uncertainty on the measured luminosity distance. This allows us to capture the correlations between the measured signal strength and the measured redshift, which is necessary in order to model the selection effects consistently between the detection model and the calculation of $P_\mathrm{det}$ (Equation~\ref{eq:beta}).

We set the ``typical'' SNR, $\rho_0$, of a BBH system with parameters $\mathcal{M}$ and $d_L$ to:
\begin{equation}
\label{eq:rho0}
\rho_0 \equiv 8 \left( \frac{\mathcal{M} \left(1+z\right)}{\mathcal{M}_8} \right)^{5/6} \frac{d_{L,8}}{d_L},
\end{equation}
where we fix $\mathcal{M}_8 = 10 \ M_\odot$ and $d_{L,8} = 1$ Gpc. This scaling approximates the amplitude of an inspiral GW signal to first order, and we chose $\mathcal{M}_8$ and $d_{L,8}$ to roughly match the typical distances of detected sources by aLIGO at design sensitivity~\citep{2017arXiv170908079C}.\footnote{It should be noted that this scaling breaks down for high-mass sources, where a significant fraction of the SNR comes from the merger and ringdown components of the signal, rather than the inspiral. However, the heaviest BBHs in our simulated population are 40--40 $M_\odot$ in the source frame, and so our synthetic model provides a good approximation to their detectability.} The true SNR, $\rho$, in our model depends on the angular factor $\Theta$, and is given by:
\begin{equation}
\rho = \rho_0 \Theta,
\end{equation}
similar to relationship between the true SNR and the ``optimal SNR'' via the projection factor $\Theta$ \citep{FinnChernoff:1993} or $w$ \citep{2015ApJ...806..263D,2017arXiv170908079C}, although $\Theta$ in our case is simply a random variable with a log-normal distribution. As the uncertainty on $\Theta$ controls the uncertainty on $d_L$, we pick the variability of $\Theta$ in order to get realistic measurement uncertainties on $d_L$. We find that realistic measurement uncertainties on $d_L$ are achieved when $\Theta$ has a typical width of 15\%, and so we pick:
\begin{equation}
\label{eq:Theta}
\log \Theta \sim N\left(0,\frac{0.3}{1+\frac{\rho_0}{8}}\right).
\end{equation}

From the true parameters $\mathcal{M}$, $\eta$, $d_L$ and $\Theta$, we assume that the measurement process measures three parameters: the observed SNR, $\rho_\mathrm{obs}$, the observed chirp mass, $\mathcal{M}_\mathrm{obs}$, and the observed symmetric mass ratio, $\eta_\mathrm{obs}$. These are given by:
\begin{align}
\rho_\mathrm{obs} &= \rho + N(0,1), \\
\log \mathcal{M}_\mathrm{obs} &= \log\left(\mathcal{M}(1+z)\right) + N\left(0,8\frac{\sigma_\mathcal{M}}{\rho_\mathrm{obs}} \right), \\
\eta_\mathrm{obs} &= \eta + N\left(0,8\frac{\sigma_\eta}{\rho_\mathrm{obs}} \right),
\end{align}
where we assume that the observed SNR is normally distributed about the true SNR, $\rho$, with a standard deviation of 1 (due to different realizations of Gaussian noise), and the uncertainties on the mass parameters scale inversely with the observed SNR~\citep{PhysRevLett.115.141101}.
We fix a threshold of $\rho_\mathrm{obs} \geq 8$ for detection. To match the expected measurement uncertainties, we fix $\sigma_\mathcal{M} = 0.04$ and $\sigma_\eta = 0.03$, so that the relative 90\% credible interval uncertainty for the recovered detector-frame primary (secondary) mass is typically 40\% (50\%)
\citep{2016PhRvD..94j4070G,2017PhRvD..95f4053V}. Meanwhile, the luminosity distance is measured from $\rho_\mathrm{obs}$ via Equations~\ref{eq:rho0}--\ref{eq:Theta}. The typical relative 90\% confidence interval uncertainty for the recovered luminosity distance is $\sim$ 50\%, which is also a realistic expectation \citep{2017PhRvD..95f4053V}. As discussed earlier in this section, this process of recovering the measured luminosity distance from the measured signal strength is necessary in order to incorporate selection effects consistently, and ensure that we generate single-event posteriors, $p(m_1, m_2, z)$ that are compatible with the assumed detection probability, $P_{\rm det}(m_1, m_2, z)$.

Under this synthetic (yet realistic) detection model, we generate 500 \emph{detected} BBH systems for each of the two populations. The projected constraints on the power-law slope and redshift evolution parameter are shown in Figure~\ref{fig:simconstraints}. Note that the maximum mass parameter will be already tightly measured with a few tens of detections \citep{2017ApJ...851L..25F}. We find that after 100 detections by LIGO-Virgo (which may happen as early as the next observing run, starting in late 2018) it may be possible to detect deviations from a uniform in comoving volume merger rate if the true redshift distribution evolves as steeply as the low-metallicity SFR ($\lambda \sim 3$). (This is expected from formation channels where the typical time delay between formation and merger is short.) Additionally, we expect to distinguish between a merger rate density that increases with increasing redshift and a merger rate density that decreases with redshift (as expected from formation channels with very long time delays). If the true deviation from a uniform merger rate density is small ($0 < \lambda < 1$, where $\lambda = 0$ implies a uniform merger rate density), it may take $\sim500$ detections to confidently exclude $\lambda = 0$. This will require a few years of aLIGO operating at design sensitivity \citep[starting in 2020+][]{Abbott:OS}. Meanwhile, extreme deviations from a constant merger rate density ($\gamma \neq 3$ in Redshift Model B) can be ruled out in 100 detections, as $\gamma$ will be constrained to a 90\% credible interval width of $\lesssim 1$. 

In summary, we expect that with $N$ detections by LIGO-Virgo operating at design sensitivity, we will be able to constrain $\lambda$ from Model A to a 90\% credible interval of width $\sim 31/\sqrt{N}$ and $\gamma$ from Model B to a 90\% credible interval of width $\sim 8.5/\sqrt{N}$, although the exact rate of convergence depends on the true values of these parameters. If the local BBH merger rate is 100 Gpc${^-}3$ yr$^{-1}$, the mass distribution follows a power law with $\alpha = 1$ and $M_\mathrm{max} = 40 \ M_\odot$, and the redshift distribution is constant in comoving volume in source-frame time, we expect $\sim 300$ detections in one year of LIGO/Virgo operating at design sensitivity (assuming that each detector has a duty cycle of 80\%, and that a confident detection requires at least two detectors in observing mode). Fixing this mass distribution, but assuming instead that the redshift distribution follows the Madau-Dickinson SFR, the expected number of detections increases to $\sim 680$. On the other hand, if the overall merger rate is the same, but the BBH mass distribution follows a steeper power law with $\alpha = 2.35$ (fixing $M_\mathrm{max} = 40 \ M_\odot$), we expect $\sim 140$ detections in a year of LIGO/Virgo operating at design sensitivity if the merger rate density is independent of redshift, and $\sim 280$ detections if the merger rate density follows the SFR. (For O3 sensitivity, the expected number of detections is smaller by a factor of 4--5.) With 100--700 detections per year, we expect to detect deviations from a constant merger rate density, or severely constrain such deviations, within the first 1--3 years of LIGO/Virgo operating at design sensitivity (starting $\sim 2020$).

\section{Discussion}
We have explored the ability of the LIGO-Virgo network to measure the redshift evolution of the BBH population.
We have applied a simple four-parameter model to constrain the BBH merger rate density, $\frac{dN}{dm_1dm_2dz}$, as a function of component masses and redshift. Our model allows us to simultaneously constrain the slope and maximum mass of the distribution of primary BH masses, the slope of the redshift distribution, and the merger rate. We note that our method can also be applied to the binary neutron star (BNS) population, although such sources will only be detectable up to redshifts $z < 0.1$ with the current generation of GW detectors. However, the mass distribution of such sources may already be well constrained from the galactic population \citep{2012ApJ...757...55O}. If we adopt this mass distribution as a prior, the analysis on BNS would simplify to a one-parameter redshift evolution model.

Recall that a measurement of the merger redshift distribution constrains a combination of the formation rate as a function of redshift and the time-delay distribution. If we can constrain the redshift evolution parameter to $\lambda \gtrsim 2.4$, we may infer that the BBH formation rate density peaks at higher redshift than the SFR, regardless of the time-delay distribution. If we assume that the time-delay distribution follows Equation~\ref{eq:td} with $\tau_\mathrm{min} = 50$ Myr and $\tau_\mathrm{max} = 14$ Gyr, a measurement of $\lambda \gtrsim 1.3$ implies that the BBH formation rate density peaks at higher redshift than the SFR. Alternatively, if we assume that the formation rate density follows the low-metallicity SFR in Equation~\ref{eq:lowZSFR}, measuring $\lambda \gtrsim 1.9$ ($\lambda \lesssim 1.9$) would allow us to infer that the time-delay distribution is more skewed towards short (long) time delays than Equation~\ref{eq:td}. 

Because our focus is on extracting the redshift evolution of the merger rate, we have simplified our treatment of the mass distribution. For example, our parametrization assumes that the BBH mass distribution does not evolve with redshift, and does not allow the distribution of mass-ratios or the minimum BH mass to vary. We have verified that adding two free parameters, $\beta$ and $M_\mathrm{min}$, to describe the mass-ratio distribution and the minimum mass according to:
\begin{equation}
p(m_1,m_2 \mid \alpha, \beta, M_\mathrm{max},M_\mathrm{min}) \propto \frac{m_1^\alpha m_2 ^\beta}{m_1-M_\mathrm{min}} \mathcal{H}\left(M_\mathrm{max}-m_1\right),
\end{equation}
does not significantly affect our results for the six BBH detections. Taking uniform priors on these additional parameters, $-4 < \beta < 4$, $3 < M_\mathrm{min} < 9 \ M_\odot$~\footnote{The minimum BH mass is constrained to be no larger than $\sim 9 \ M_\odot$, the 95\% upper bound on the secondary mass of GW170608~\citep{2017ApJ...851L..35A}.},  causes the posteriors on the remaining four parameters to widen only slightly. Furthermore, the constraints on the additional two parameters, $\beta$ and $M_\mathrm{min}$, are not informative with only six detections, and are consistent with our default values, $\beta = 0$, $M_\mathrm{min} = 5 \ M_\odot$, although we find a slight preference for equal mass ratios ($\beta > 0$ at $\sim70\%$ credibility; consistent with the results of~\citealt{2018arXiv180610610R}) and a larger minimum mass ($M_\mathrm{min} > 3.9 \ M_\odot$ at $95\%$ credibility). These results hold for both Models A and B of the redshift evolution. It will likely take $\mathcal{O}(100)$ detections to measure $M_\mathrm{min}$ sufficiently well and resolve the putative gap between the neutron star and BH mass spectrum~\citep{Littenberg:2015,Mandel:2017,Kovetz:2017}. 

With sufficient detections, our four-parameter model will likely break down, and we should include more degrees of freedom in the mass-redshift (and possibly also spin) distribution to avoid introducing systematic biases in the inferred parameters \citep{2018ApJ...856..173T,Wysocki:2018}. \maya{For example, if the mass distribution varies with redshift, possibly favoring larger masses at high redshifts due to the lower average metallicity, our simple model will misinterpret this as an evolution in the merger rate. Therefore, a more complicated model should allow for correlations between the mass and redshift distribution, either through the addition of one to two parameters (e.g. a copula model), or a many-parameter model that fits the mass distribution separately in different redshift bins. It may also be possible to introduce a multi-component mixture model to determine whether there are multiple populations of BBHs following different mass-redshift distributions. Furthermore, a more sophisticated model would include at least two additional parameters to fit the peak, $z_\mathrm{peak}$, and the high-redshift slope of the merger rate density. For example, we can consider the following parametrization of the merger rate density inspired by the Madau-Dickinson SFR:
\begin{equation}
\begin{aligned}
p(z \mid a, b, z_\mathrm{peak}) &\propto \frac{1}{1+z}\frac{dV_c}{dz}\frac{(1+z)^a}{1+\left(\frac{a}{b-a}\right)\left[\frac{1+z}{1+z_\mathrm{peak}}\right]^b}.
\end{aligned}
\end{equation}
This parametrization provides an excellent fit to all of the astrophysical redshift distributions discussed in Section~\ref{sec:model} up to redshifts $z \sim 4$, whereas our one-parameter model of Equation~\ref{eq:uc_dev} starts to break down at $z \sim 1$. However, it is unlikely that we will have tight constraints on $z_\mathrm{peak}$ and $b$ with second-generation GW detectors, because $z_\mathrm{peak}$ will likely lie beyond the sensitivity of these detectors, unless the time delays between formation and merger are typically extremely long (greater than a few Gyr). For example, if the BBH formation rate follows the Madau-Dickinson SFR and the time-delay distribution in Equation~\ref{eq:td}, the peak of the merger rate density would be at $z_\mathrm{peak} = 1.4$, where we expect to have very few detections (see Figures~\ref{fig:cumpzdet} and \ref{fig:pzdet}). If the BBH formation rate peaks later than the Madau-Dickinson SFR, following the low-metallicity SFR in Equation~\ref{eq:lowZSFR} for example, the peak of the merger rate density would be even farther out of reach, at $z_\mathrm{peak} = 2.1$, assuming the same time-delay distribution. However, by the time we have $\mathcal{O}(1000)$ detections, it may be possible to get some measurement of $z_\mathrm{peak}$, which would allow us to infer the peak of the formation rate density for an assumed time-delay distribution (this peak would be at $z = 1.8$ if BBH formation followed the Madau-Dickinson SFR, and $z=2.7$ if the formation followed the low-metallicity SFR of Equation~\ref{eq:lowZSFR}). We look forward to a time where our single redshift-evolution parameter is sufficiently well measured that we must include these additional parameters in our model; we anticipate that this will take over 500 detections. We note that with third-generation GW detectors, it will be possible to accurately infer the entire formation rate history of BBHs together with the time-delay distribution from the observed redshift evolution of the merger rate \citep{2018arXiv180800901V}.}

In addition to the limitations of our parametrized model, another much less significant source of systematic uncertainty in our analysis comes from GW measurements of the luminosity distance (and therefore, the redshift) to a source. Extracting the luminosity distance from a GW signal depends on measuring its amplitude, which is affected by detector calibration uncertainties. The calibration uncertainty is only a few percent~\citep{2016RScI...87k4503K}, which is negligible compared to the expected uncertainty on the redshift evolution parameter. Another subdominant source of uncertainty comes from the effect of weak lensing on the GW amplitude, which contributes at the sub-percent level and is therefore negligible for our analysis~\citep{1998PhRvD..58f3501H,2005ApJ...631..678H}.

\section{Conclusion}
By fitting a four-parameter mass-redshift distribution to the first six announced BBH mergers, we have placed the first constraints on the redshift evolution of the BBH merger rate. We show that because of strong correlations between the masses and redshifts of detected BBHs, the mass and redshift distribution must be fit simultaneously. We consider two parametrizations of the redshift evolution: Model A fixes the slope of the redshift distribution to match the differential comoving volume locally (as $z \to 0$), as is expected from most astrophysical formation channels, while Model B allows for large deviations, even at low redshift. Our constraints from six events are too weak to distinguish between any astrophysical formation scenarios: for example, we measure $-31 < \lambda < 5$ at 90\% credibility for Model A, whereas typical formation channels predict between $-4\lesssim\lambda\lesssim3$). However, we can already constrain extreme deviations from a uniform in comoving volume merger rate, finding $\gamma = 2.7^{+1.8}_{-1.3}$ for Model B ($\gamma = 3$ corresponds to a uniform merger rate). 
Furthermore, while previous analyses have calculated the BBH merger rate under the assumption of a uniform in comoving volume redshift distribution, we demonstrate how to infer the merger rate density as a function of redshift. 

We project that with 100--500 detections by LIGO-Virgo, the inferred redshift evolution of the BBH merger rate will allow us to distinguish between proposed formation channels (for example, those that favor long versus short time delays between progenitor formation and BBH merger). Meanwhile, the overall merger rate and mass distribution will also provide important clues regarding the formation channel \citep{2015ApJ...806..263D,2015ApJ...810...58S,2017ApJ...846...82Z,2018MNRAS.477.4685B}. In 5--10 years, the constraints on the redshift evolution parameter alone will allow us to infer the peak of the formation rate of BBH progenitors and/or the typical time delay between formation and merger.

\acknowledgments
We thank Christopher Berry, Eve Chase, Reed Essick, and Peter Shawhan for their helpful comments on the manuscript. MF was supported by the NSF Graduate Research Fellowship Program under grant DGE-1746045. MF and DEH were supported by  NSF grant PHY-1708081. They were also supported by the Kavli Institute for Cosmological Physics at the University of Chicago through NSF grant PHY-1125897 and an endowment from the Kavli Foundation. DEH also gratefully acknowledges support from the Marion and Stuart Rice Award. WMF is supported in part by the STFC.

\bibliographystyle{yahapj}
\bibliography{references}

\end{document}